\documentclass[a4paper,twoside]{article}

\usepackage{epsfig}
\usepackage{subcaption}
\usepackage{calc}
\usepackage{amssymb}
\usepackage{amstext}
\usepackage{amsmath}
\usepackage{amsthm}
\usepackage{multicol}
\usepackage{pslatex}
\usepackage{apalike}
\usepackage{algorithm2e}
\usepackage[bottom]{footmisc}
\usepackage{booktabs}
\usepackage{graphicx}
\usepackage{color}
\usepackage{xltabular}

\usepackage{SCITEPRESS}     

\begin{document}

\title{Students' Perception of Big Data Engineering in Higher Education Curricula: Expectations, Interest and Ethical Implications}

\author{\authorname{Ioana-Georgiana Ciuciu\sup{1}\orcidAuthor{https://orcid.org/0000-0002-7126-0585}, Manuela-Andreea Petrescu\sup{1}\orcidAuthor{https://orcid.org/0000-0002-9537-1466} }
\affiliation{\sup{1}Faculty of Mathematics and Computer Science, Babe\c{s}-Bolyai University, Cluj-Napoca,
Romania}
\email{\{ioana.ciuciu, manuela.petrescu\}@ubbcluj.ro}
}

\keywords{Big Data Engineering, Emerging Technologies, Higher Education, Curricula, Learning Path, Learner Empowerment, Ethics, Collaborative and Interdisciplinary Approaches to Learning.}

\abstract{The study investigates students' interest and expectations in a Big Data Engineering course integrated with a Master curricula, as well as ethical implications of using Big Data. An anonymous online survey was conducted with 42 of the 67 students enrolled in the Big Data course offered to Computer Science and Bioinformatics Master's programs. The responses were analyzed and interpreted using thematic analysis, highlighting interesting aspects related to students' expectations, interest, and their perspective of the ethical implications of working with Big Data. The study concludes that, even though there is significant difference in students' background, the majority are interested in learning Big Data, for practical and personal reasons related to the potential for career growth and their passion for the field. The main expectation expressed is related to enhancing their knowledge related to Big Data via practical activities. All students demonstrate awareness of potential ethical threats related to security and privacy, while Computer Science students are aware of the possibility of introducing bias in data during acquisition and analysis and of potential abusive data usage.}

\onecolumn \maketitle \normalsize \setcounter{footnote}{0} \vfill

\section{\uppercase{Introduction}}
If we look at the perceived value of Big Data today, we can notice that Big Data, far from being a hype any more, is prevalent in our daily lives throughout many hype cycles\footnote{https://www.gartner.com/en, last accessed on 16.02.2025}. Moreover, Big Data is an integral part of the 4th Industrial Revolution, being used to offer solutions to real-life problems in almost every possible domain. The concept is illustrated in \cite{Mamadou} with a parallel between the economic and industrial digitalization known as Industry 4.0 and the evolution of higher educational institutions in order to adapt and respond to the specific needs of learners, introduced as University 4.0. Integrating Software Engineering with emerging technologies of the present such as Artificial Intelligence (AI), the Internet of Things (IoT) and Cloud Computing, the field of Big Data Engineering allows implementing data ecosystems that make possible the handling (i.e., storage and processing) of massive amounts of heterogeneous data generated in real-time. Most importantly, it offers the infrastructure to store, process, and visualize data that is then served to various applications for analysis and decision-making. 

In this context, it is important that higher education programs prepare students to reach their full potential in the domain of Big Data Engineering by the time of graduation. This is only possible by exposing them early enough to theoretical and applied knowledge from these emerging domains, through dedicated and well-aligned curricula \cite{Kim}. Students should be able to experiment with emerging data engineering technologies via collaborative, hands-on projects with real-world interdisciplinary case studies and continuous industry-academia collaborations, in order to build strong competencies and prepare them for their future careers. 

This study investigates the transformative power of the Big Data technologies when included in the Master program curricula, from the students' perspective. We aim to empower students to reach their full potential and desired careers in Big Data through dedicated, collaborative, interdisciplinary learning.  Our work contributes to the discussion of best practices in software engineering education for emerging technologies, specifically Big Data.

We addressed the particular case of the students enrolled in a Big Data course, titled '\textit{Big Data Processing and Applications}', given in English at Master level from the Faculty of Mathematics and Computer Science, Babes-Bolyai University, Cluj-Napoca, Romania. 
The main objective of the study is to illustrate students' expectations related to Big Data and their interest in following a career path related to Big Data by analyzing factors beyond their attraction and motivation in Big Data technologies. A secondary objective of the study is to analyze ethical issues in Big Data. 
The objectives of our study have been divided into the following research questions that address the students enrolled in the Master Programs in Computer Science and Bioinformatics at Babe\c{s}-Bolyai University after completing two semesters:

\textit{\textbf{RQ1.} What is the students' experience in IT?}

\textit{\textbf{RQ2.} What are the students' expectations from the Big Data course?}

\textit{\textbf{RQ3.} What is the students' interest related to Big Data}

\textit{\textbf{RQ4.} What are the ethical issues in students' opinion? Does background influence it?}

The rest of the paper is organized as follows: Section \ref{litreview} describes related work, Section \ref{studydesign} addresses the research methodology of the present study, Section \ref{results} presents the results of the thematic analysis, while Section \ref{discussion} interprets the results and highlights important aspects with respect to the identified research questions. Section \ref{threats} summarizes the potential threats to validity and the actions taken to minimize and mitigate them. Section \ref{conclusion} presents our conclusion and future work.

\section{\uppercase{Literature Review}} \label{litreview}

Digitalization and emerging technologies demand skilled professionals to manage real-world Data Science projects and adapt to the evolving technologies landscape. Past and recent studies have shown the need to bridge the gap between university curricula and employment regarding emerging technologies, and in particular Big Data. 
A 2014 study \cite{Sigman} identified a significant demand for employees who are proficient in Big Data technologies. The study emphasized the role of universities in fulfilling this demand and proposed collaborative initiatives between academia and industry to optimize Big Data education. 

In a research survey from the US by \cite{Song}, titled "Big Data and Data Science: What should we teach?", the authors identified the biggest bottleneck in the Big Data era as the production of capable data scientists. The authors conclude that technologies, tools, and languages can be learned, but professionals who can do real-world Data Science projects and possess the necessary Big Data skills and knowledge are rare. Regarding the role of digitalization in improving the skills and employability of engineering students, the authors in \cite{Xu} identified that both hard and soft skills are important in Big Data curricula, the latest having a greater impact on employability. Students need to be taught fundamental concepts, algorithms, and architectures, but most importantly they need to have hands-on experience with Big Data technologies.

A relatively recent study by \cite{Baig} reviewed the main research directions related to Big Data in education.
The integration of Big Data into the curriculum was the least represented direction in the literature (only 10\% of the reviewed studies). The results show that integrating Big Data into the curriculum requires an adaptation of the curriculum to the learning environment and a good alignment between the training factor, learning objectives, and results, also confirmed by other studies such as \cite{Nelson,Sledgianowski}. The authors of these studies discuss the importance for teachers to be able to measure students learning behaviour and attitude simultaneously and adapt their teaching strategy accordingly. 
The integration of Big Data into the curriculum has been a challenge in many fields of education \cite{Dzuranin}. Big Data analysis also integrates generative artificial intelligence (i.e., Chat GPT). 
In \cite{Valle}, the authors analyze the students' behavioral intention toward using Artificial Intelligence in education. The results show that social influence and perceived knowledge of the use of AI are significant predictors of the attitude toward the use of AI. 
In \cite{Kumar24} authors provide a comprehensive examination of the positive and negative impacts of generative AI on higher education.

Similarly, in this study we propose to investigate students' attitude towards learning and using Big Data by analyzing their interest and expectations in this field. The aim is to continuously improve and adapt the teaching strategy in a collaborative and dynamic setting where students practice research-informed hands-on project tasks within multi-disciplinary project teams around real-world use cases, in tight relation with the IT industry.

Research on the ethical implications of Big Data and AI highlights the challenges and potential impact on society in general and higher education in particular \cite{Komljenovic,Lundie}. Researchers emphasize the critical need to integrate the ethical frameworks of AI and Big Data into higher education \cite{Kumar24}. By exploring student awareness of ethical considerations related to Big Data in this paper, we aim to contribute to the critical aspect of responsible data handling and the ethical implications of its use within higher education.

\section{\uppercase{Study Design}} \label{studydesign}
The study's design and data gathering details are addressed in this section. This study's design complies with recognized community guidelines for qualitative surveys (Ralph, Paul (ed.), 2021). The following pillars served as the foundation for the study:\\
\textit{Scope}: The main scope of this paper was to find the student's interest and expectations, as well as ethical implications of big data. \\
\textit{Period}: The study was based on an anonymous online survey that remained open for the first two weeks, at the beginning of the course.\\ 
\textit{Method}: We used a hybrid strategy for data analysis. The methodology integrated qualitative analysis of the data collected for answers to open-ended questions with quantitative analysis for answers to closed questions.\\
\textit{ Tools}: During the data analysis process, we stored intermediate data using various tools and we used Microsoft Teams for the survey because students were already enrolled in the platform. 

\subsection {Survey Design}
\label{sec:interviewDesign}
With a focus on the study's research scope, each author elaborated a set of survey questions, both authors analyzed the set of questions, discussed and agreed on which questions to ask, and also agreed on the final version of each question. 
Although closed questions allowed us to classify and group responses, open-ended questions were designed to allow for a more thorough examination of the student's interests and point of view.
The closed questions -  Q1, Q2, and Q3- consisted of questions related to the student's specificity, background, and work experience in the IT domain. We did not ask for information that could help us identify the participants. The open questions were related to Big Data, the questions are listed in Table\ref{tab:questions}.

\begin{table}[h!]
  \caption{Survey Questions}
  \label{tab:questionnarie}
  {\small{
  \begin{tabular}{p{7cm}}
\toprule
Q1. Please specify your domain of study. \\
Q2. Do you work in the IT industry? \\
Q3. Please specify for how long you have worked (years).\\
Q4. What are your expectations related to this Big Data course?\\
Q5. What knowledge related to AI / Big Data do you have? \\
Q6. Why are you interested in the field of Big Data? \\
Q7. What do you consider to be the main ethic-related issues when dealing with Big Data? \\
   \bottomrule
\end{tabular}
}}
\label{tab:questions}
\end{table}
Even a short period of work experience, as students might have (internships, short-term contracts), provides valuable practical insights that help deepen one's understanding of a domain, so we added work experience-related questions (Q2 and Q3).

\subsection{Participants}
Our survey's target audience was formed by second-year Master students studying Bioinformatics (from the Faculty of Biology) and Computer Science (from the Faculty of Mathematics and Computer Science), enrolled in the Big Data course. The set of participants consisted of 67 students, of whom 42 consented to participate in the study. There was no participant selection; everyone enrolled in the course was informed and could take part in the survey. We paid attention to ethics, so we informed the students that the survey was anonymous and the participation was voluntary. We also let them know how we are going to use the collected data. 

We identified and analyzed the two groups (Bioinformatics (BI) and Computer Science (CS)) of students' responses on an interdisciplinary basis. When we started analyzing the data, we noticed that there was a high degree of similarity in the answers received from the students enrolled in Computer Science different lines of study. For this reason, we considered that a comparative analysis of the responses received from CS students is not relevant and does not represent a scientific interest and that an interdisciplinary point of view would be more interesting. The cohorts of CS students were also more similar in terms of work experience: most of them had at least one internship experience in different companies, compared to the BI group, where only one person had IT working experience. In addition, having too many groups for analysis could affect the presentation of the paper by increasing the complexity without adding a scientific benefit.
In conclusion, we consider that a comparative analysis of the results between the two groups, BI and CS, is more relevant for the purpose of this study.

\subsection{Methodology}

We conducted a survey with open-ended and closed-ended questions. Quantitative methods were used to analyze and interpret the answers to open-ended questions. The questionnaire surveys were in accordance with the established standards of the empirical community \cite{ACM}.

\textbf{Data Collection.} When we proposed that the questions be set in English, we considered two aspects: the first was that their field of study is English-based. The second aspect was to reduce any potential risks associated with translation. The responses' data were gathered just as they were written by the students, without any alterations. The survey remained open for two weeks. The anonymous survey link was sent through their MSTeams faculty account.

\textbf{Data Analysis.} We used thematic analysis to interpret the responses received for the open questions and performed the following next steps:

\begin{itemize}
    \item [] 
    \begin{enumerate}
        \item The authors independently attempted to find codes or key items within the responses received.
        \item The resulting key items were grouped into themes according to their frequency using techniques such as generalization, removal, and reassignment to move the key items with low occurrence to larger themes.
        \item The final phase involved a discussion among the authors related to supporting information for the categorization process, representations, and other issues.
    \end{enumerate}

\end{itemize}

Although some students' comments were brief, many others contained up to four or five statements or explanations (which could be categorized into 4-5 key items). A response may therefore include one or more key items; as a consequence, the sum of all percentages in our analyzes will be greater than 100\%.

\section{\uppercase{Results}} \label{results}
The subsequent analysis focuses on the evaluation of the student responses to the survey in relation to the four main research questions of the study. Each question enables a comprehensive examination of the perceptions, interest, expectations, and ethical issues of the students based on the responses received. The data collected are comparatively analyzed based on students' background, knowledge, and experience. 

\subsection{RQ1. What is the students' experience in IT?}

To find the answer to this question, we analyzed the responses received for questions Q1, Q2, and Q3 of the survey.
The first question was needed as there were students from five lines of study enrolled in this course, and we expected to have differences in their responses based on the line of study, as follows: Bioinformatics - 13 students; Data Science - 20 students; High Performance Computing - 12 students; Software Engineering - 20 students; and Applied Computational Intelligence - 2 students. 

\begin{figure}[!htbp]
    \centering
    \includegraphics[width=1\linewidth]{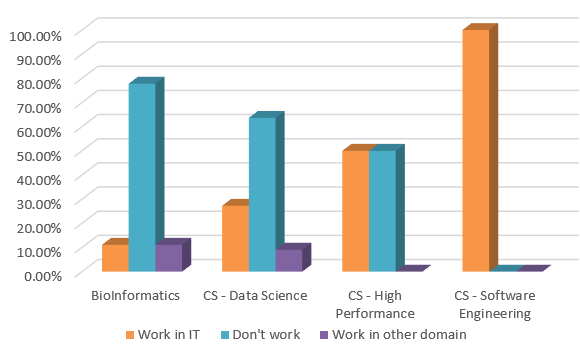}
    \caption{Students' distribution by line of study and work experience}
    \label{fig:WorkBio}
%   \vspace{-5mm}
\end{figure}

From second question answers, we found out that only one of the students from Bioinformatics works in an IT related domain. Students in Computer Science have the majority of jobs in IT related fields. For example, most of the students in Software Engineering work in IT, as can be seen in Figure \ref{fig:WorkBio}. The values in the figure represent the percentages of students who work (in any domain) compared to the number of students enrolled in a specific line of study.

\begin{figure}[!htbp]
    \centering
    \includegraphics[width=1\linewidth]{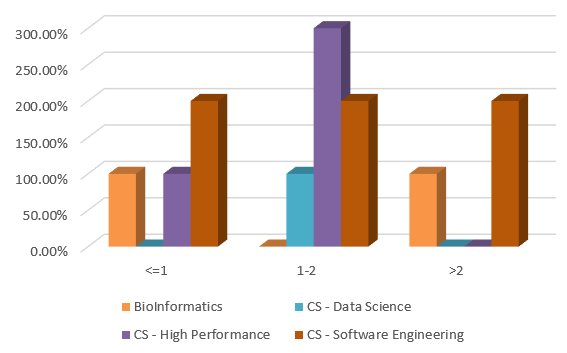}
    \caption{Students' distribution by years of work experience in IT}
    \label{fig:WorkYears}
%   \vspace{-5mm}
\end{figure}

As we considered the IT work experience to be relevant, even if students might not have work experience in Big Data, but have overall programming experience, so we asked for the number of worked years in the IT domain. Most of them (with a few exceptions) have up to 2 years of work experience (Figure \ref{fig:WorkYears}).

Based on these answers, students from the Computer Science lines of study seem to have more in common as compared to students enrolled in Bioinformatics. As there are similarities between Computer Science enrolled students, we decided to focus in this research on two groups: Bioinformatics and Computer Science students.

\subsection{RQ2. What are the students' expectations from the Big Data course?}

To find the answer to this question, we analyzed the responses received in the survey questions Q4 and Q5. The purpose was not only to see their expectations but also to see if there is a correlation between their experience and their expectations.

\begin{figure}[!htbp]
    \centering
    \includegraphics[width=1\linewidth]{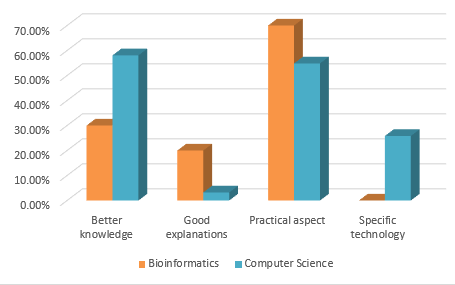}
    \caption{Students' expectations from the Big Data course}
    \label{fig:Expectations}
%   \vspace{-5mm}
\end{figure}

One student did not reply to this question, all the other specified at least one thing they would like to learn. When we analyzed the data, we took into consideration all the ideas/ key items mentioned by students, and we computed the percentages using the following ratio: number of appearances versus the total number of students in the specific cohort (Bioinformatics or CS). As most of the students mentioned more than one, the sum of the percentages of all key elements exceeds 100\%. The same method of calculus will be used for all the percentages in the figures from Subsections RQ2, RQ3 and RQ4. The students enrolled in CS had more expectations and mentioned more key items compared to Bioinformatics enrolled students, but both cohorts scored high in practical aspects, in collecting, storing, and using Big Data. The main classes of expectations differentiated by study line are presented in Figure \ref{fig:Expectations}.

Practical aspects represent the most mentioned objective and many times are correlated with the desire to improve their knowledge: \textit{''Find some knowledge about big data and situations when it is useful and how to use it.'', ''Learn how the data is gathered and stored.''}. The interest in learning more is sometimes related to economical aspects and to career: \textit{''I want to learn more about this subject and understand how can I apply it in my career''}.

The second most mentioned objective is improving their knowledge: \textit{''learn new ways to handle big data in general'', ''to improve my skills''}. However, some students seem to be aware of the differences between the students from Computer Science and those from Bioinformatics. CS students have a different set of skills and IT related knowledge and it is difficult to cope with their learning pace: \textit{''I do not personally like the idea of multidisciplinary in this course. In the past, it has not worked too well. People from Bioinformatics have their own rhythm''}. 

Computer Science students were a little bit more specific and knowledgeable in what they want to learn, proving that they already have some basic knowledge or concepts. They mentioned: \textit{''Data pipelines and some automation'', ''improve my knowledge of Big Data, the architecture and the analysis of it''}.

Based on the key items found, the students' knowledge oscillates between none/hobby level, to basic, to intermediate: \textit{''almost none'', ''I think that my knowledge in this domain is at a basic to intermediate level''}. % - as can be visualized in Figure 
%\ref{fig:CurrentKnowledge}.
Some students mention that they have basic knowledge: \textit{''Basic knowledge: visualization, simple algorithms (clustering, classification, decision tree...), data cleaning, dashboard''}, others just state that they \textit{''work with Big Data and AI''}. However, based on the results, Bioinformatics students selected this course based mainly on personal interest and have less knowledge in the domain compared to Computer Science students who have basic knowledge in Big Data and have previously had AI courses \textit{''I studied AI during University and Master courses and did my bachelor’s degree in this domain''}.

In conclusion, all students want to improve their knowledge, even if the base level is different between the two groups, Computer Science and Bioinformatics students. The previous knowledge of AI/Big Data is correlated with the line of study, suggesting that even if there was a personal interest in AI/Big Data, students do not learn by themselves. All students express a high interest in practical aspects of data gathering, data storage, and data manipulation.

\subsection{RQ3. What is the students' interest related to Big Data?}

\begin{figure}[!htbp]
    \centering
    \includegraphics[width=1\linewidth]{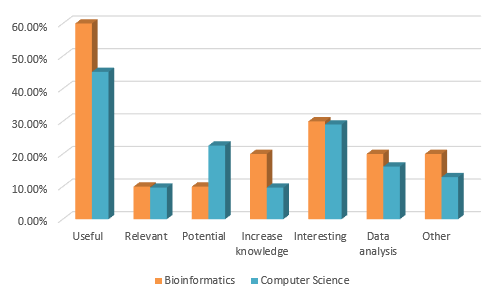}
    \caption{Students' interest related to Big Data}
    \label{fig:Interests}
%   \vspace{-5mm}
\end{figure}

After analyzing the responses received to the survey question Q6 ("Why are you interested in the field of Big Data?"), the practical aspects came through. We grouped the selected items in two categories: \textbf{Practical} reasons that includes ''Useful'', ''Relevant'' and having ''Potential'' domain of activity and \textbf{Personal} reasons which includes ''Interesting'', ''Increase knowledge'', to know how to do ''Data Analysis". Other reasons such as \textit{''how to keep the data clean and secure''} were less prevalent and were grouped into ''Other'' reasons as can be seen in Figure \ref{fig:Interests}.

All students are aware of the potential and usefulness of Big Data, regardless of their background. They perceive the importance of manipulating and getting information from large data sets, in Bioinformatics \textit{''because Bioinformatics deals with large amount of data, and I would like to know how to process them''} and Computer Science \textit{''I'm particularly interested in data analysis, and extract capital information from big batch of data''}. Being able to extract results is seen as a potential solution for problems humans find difficult to solve: \textit{''I’m curious about how data can be used to uncover insights and solve complex problems'', ''I am interested in Big Data because I believe it has the potential to drive innovation and improve decision-making in various fields, from healthcare to business''}. Sometimes they refer to the general impact: \textit{''the world basically revolves around large amounts of data being stored and processed''}, others are more focused on their own benefit: \textit{''Because I enjoy working with data and using it is a crucial step in my industry (finance and trading)''}. However, the largest majority of students (from both sets- Bioinformatics and Computer Science) appreciate the usefulness of the domain as the major factor for their interest in Big Data. 

A smaller subset of the respondents pointed out their own preferences related to the domain: \textit{''it's a very vast and interesting domain''} or to their feelings: \textit{''I am curious how is it to work with Big Data''}.

In conclusion, the students' interest related to Big Data lay mainly in practical aspects: usefulness and its potential for growth (implying more, better paid jobs, more positive impact for society). For a smaller percentage, an important factor are personal passions.

\subsection{RQ4. What are the ethical issues in students'
opinion? Does background influence it?}

Based on the answers received to survey question Q7, we found a set of key items related to the ethical concerns of students, and grouped them into the following themes:
\begin{itemize}
    \item  Security concerns: data privacy, data security, ownership, consent to use personal data;
    \item Data collection: introduced bias, collection process;
    \item Other: ecological issues, destructive potential (manipulate people);
    \item No answer / unrelated answer.
\end{itemize}

\begin{figure}[!htbp]
    \centering
    \includegraphics[width=1\linewidth]{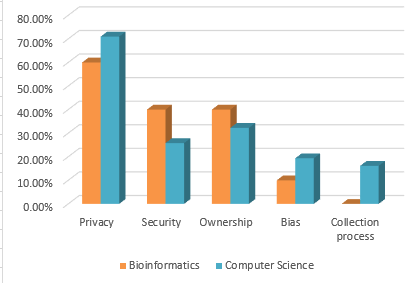}
    \caption{Students' ethical concerns related to Big Data}
    \label{fig:Ethics}
%   \vspace{-5mm}
\end{figure}

The most mentioned ethical issues were raised by all students, their lines of study did not influence the results, as can be seen in Figure \ref{fig:Ethics}.

Some students mentioned more than one concern related to data security: \textit{''data security, transparency, data ownership'', ''the possibility of personal data breach''}. Security includes (sensitive) data storage: \textit{''The main issue I consider is the careful storage of sensitive data''}, data privacy, or the use of it without the consent of the owner: \textit{''The main ethical issues in Big Data include privacy, consent, and security. Privacy is a concern when personal data is used without consent, and bias in algorithms can lead to unfair outcomes'', ''confidentiality issues especially if we think of the medical field''}. There is a difference between the Bioinformatics and Computer Science students, as the latter are more aware that the collection process and data processing could introduce bias that would mislead and influence the results: \textit{''potential bias in data collection and analysis''}.

Other ethical issues were mentioned only by Computer Science students. The ecological impact was mentioned by 9.68\% of them: \textit{''There is also ecological problem because of the use of huge data-centers''}. The scope and use of data were mentioned by other 6.45\%: \textit{''Data can be used against people to manipulate them or to abuse them''}.
We also got some unrelated answers and students that did not answer this question (6.45\% of Computer Science students) and only one from Bioinformatics students.

In conclusion, the main ethical concerns related to Big Data are relatively similar for all students, but only Computer Science students mentioned ecological or manipulative possible usage.

\section{Discussion} \label{discussion}

While examining the students' backgrounds in our participant set, we noticed a difference between their work experience, as just one of the students from Bioinformatics has work experience in IT while all the students from Software Engineering have work experience in IT. The percentages for the other groups of Computer Science students were between these extremes. There were some exceptions, students that have work experience in other domains, but their prevalence was small in our participant set. 

Based on the results in RQ1, Computer Science students seem to find more job opportunities and are already involved in the job market. 
However, most of the students are interested in practical aspects of learning Big Data, as it is perceived to be useful, has a lot of potential, and helps them perform at their jobs. Practical aspects tend to be the main reasons for choosing to enroll in this course. (Note that the course is mandatory only for students in the High Performance Computing Master Program, while for all the others, it is optional). Even if the course is perceived to be useful, practical and interesting, most of the students have no previous experience in Big Data or have mainly an experience due to the courses in their Bachelor and/or Master program. The Internet is abundant in public courses and learning resources; however, the students do not seem to have sufficiently developed their skills to learn Big Data technologies by themselves, without a professional to lead their learning process. The main question is how can we, as higher education professionals, help students learn the autodidactic skills related to Big Data? How can we prepare students to become more efficient in learning by themselves when they are interested in a specific Big Data topic?

The results on RQ2 and RQ3 confirm and align with the results from the related work of the paper, regarding the importance of practical Big Data projects, of team work and collaborative learning involving students with various levels of technical background and the continuous academia-industry collaboration in order to optimize the students' learning path and to better prepare them for their future jobs. The responses to the survey also confirm that students are aware of what is highlighted in the literature, namely the urgent need for professionals who possess the skills to learn new technologies and adapt to any Big Data project.

The results from RQ4 on ethical issues show that while all students are well aware of the main ethical challenges (with small variations based on the students' line of study), only Computer Science students mentioned possible non-ethical (manipulative) usage of data.

\section{Threats to Validity} \label{threats}
Our goal was to reduce any possible risks by analyzing and implementing the community standards described in \cite{ACM}.
Furthermore, we addressed the possible validity risks that have been identified in software engineering research \cite{ACM}.
Three aspects have been identified and examined: construct validity, internal validity, and external validity as a result of the guidelines taken into account. We paid particular attention to the participant related threats: participant set, participant selection, dropout contingency measures, and to author biases in order to ensure internal validity.

\textbf{Construct Validity:} The questions were created using a multistep procedure described in the Survey Design to reduce the writers' biases. The proposed survey questions were in line with the stated study objectives of the Introduction. 

\textbf{External validity:} We investigate the possibility of extrapolating our study's results. We point out that since we looked at a particular cohort, no generalizations can be made to the entire society. Because of this, we can, with some caution, extrapolate the results to the group of students enrolled in fields related to IT/Computer Science.

\textbf{Internal Validity:} Based on \cite{ACM}, we determined the selection of participants and participants, the dropout rates, the subjectivity of the authors and the ethics as possible internal threats. All participants in the Big Data course have been informed about the voluntary survey and invited to participate. As a result, the target group of participants was all inclusive, removing any possible dangers related to the participant selection or set of participants. Because the survey was voluntary, we had few options to reduce dropout rates. We have taken into account and investigated the potential subjectivity of the author in data processing. By employing text analysis in accordance with the recommended data processing standards, we aimed to reduce this risk. We showed participants that we were committed to ethics by informing them of our goal, by using anonymous data gathering methods, and letting them know how we intended to use the data.

\section{\uppercase{Conclusions and future work}} \label{conclusion}
The study is founded on four central research questions which we have methodologically approached to investigate students' perception of Big Data in higher education. The results confirm the main challenges related to Big Data in education, as identified in the related work in the paper. As the first implication of the study's results, students need methodological guidance from university educators in order to be able to learn and adapt to the emergent nature of the Big Data domain. Most importantly, in the field of Big Data, students need curricula that align with their learning objectives and expectations and with their future professional careers. The curricula should build competencies around fundamental concepts and techniques, and most importantly around hands-on activities. Students learn best in relation to each other; therefore, we recommend that learning occur in groups or teams of students with diverse backgrounds and experiences, in settings as close to the real world as possible. Another important factor to the success of students' learning path is the continuous academia-industry collaboration. Integrating guest lectures from industry professionals and practical workshops led by experts working daily with emerging Big Data technologies, where students can validate their competencies, is crucial for learning in this domain.

A longitudinal study to reveal further insights into the course's lasting impact is in progress. Moreover, it would be interesting to analyze how students' perceptions and interests evolve after the course. This would imply another survey, possibly combined with student evaluation results. The future approach could also be used to assess learning outcomes and their correlation with students' professional trajectories upon graduation. This would bring an important contribution towards the continuous improvement of the existing curriculum and the teaching approach, in order to better prepare students for the challenges and opportunities of Big Data Engineering.

\bibliographystyle{apalike}
{\small
\bibliography{main}}

\end{document}